# Voice Conversion Using Coefficient Mapping and Neural Network

Agbolade Olaide Ayodeji
Department of Electrical and Electronics Engineering
Federal University of Technology
Akure, Nigeria
olaideagbolade@gmail.com

Oyetunji S.A
Department of Electrical and Electronics Engineering
Federal University of Technology
Akure, Nigeria
Samlove98ng@yahoo.com

*Abstract*— The research presents a voice conversion model using coefficient mapping and neural network. Most previous works on parametric speech synthesis did not account for losses in spectral details causing over smoothing and invariably, an appreciable deviation of the converted speech from the targeted speaker. An improved model that uses both linear predictive coding (LPC) and line spectral frequency (LSF) coefficients to parametrize the source speech signal was developed in this work to reveal the effect of over-smoothing. Non-linear mapping ability of neural network was employed in mapping the source speech vectors into the acoustic vector space of the target. Training LPC coefficients with neural network yielded a poor result due to the instability of the LPC filter poles. The LPC coefficients were converted to line spectral frequency coefficients before been trained with a 3-layer neural network. The algorithm was tested with noisy data with the result evaluated using Mel-Cepstral Distance measurement. Cepstral distance evaluation shows a 35.7 percent reduction in the spectral distance between the target and the converted speech.

## I. INTRODUCTION

Voice conversion refers to a voice manipulation technique that makes words uttered by one person (source) sound like another (target). Applications of such systems are found in text to speech synthesis system personification, speaker variability reduction in voice-over, voice building, and gaming. Concatenative and parametric syntheses are the two commonly used techniques in voice conversion application. In concatenative speech synthesis, short speech segments are combined to create new speeches that contain the voice characteristics of the target speaker. These utterances which normally would contain all the phonemes present in the targeted language are decomposed into short frames of about 20 milliseconds duration to form a large corpus. New words and sentences can then be created by combining strings of individual phonemes that make up that word using certain well- defined lexical rules. This process is a fairly tedious one. First, large training samples are needed to ensure that new voices can be successfully built. There is also the challenge of having a smooth transition from one frame to another without creating a rollover effect. Thirdly, it is very difficult to adapt one voice to another. Some have even argued that concatenative speech synthesis should not be described as voice conversion since the approach merely combines phonemes to build new words with the same voice. Parametric speech synthesis system is another method used in voice conversion. This parametric approach only uses the parameters of the speech rather than the speech itself. This is a more bandwidth efficient method of converting voices. Far less training samples are required, higher flexibility is possible but more work is needed to ensure the naturalness of the synthesized speech. The efficiency of the parametric voice conversion hinges on the robustness of the speech parameterization and speech training technique. The speech parameterization refers to the technique used in representing the speech signal while the training has to do with the intelligence built into the system. Some of the popular parameterization methods are linear predictive coding (LPC), line spectral pair (LSP), cepstrum, and Mel-Frequency Cepstral Coefficient (MFCC). The LPC assumes the vocal tract functions like a physiological filter [1]. Using the autoregressive model, the speech output is represented by (1).

$$S(n) = \sum_{n=1}^{p} a_k s(n-k) + Ge(n) \qquad (1)$$

where $a_k$ is the speech coefficient, $G$ the predictor gain and $e(n)$ the excitation.

Another parametric approach is the cepstral based approach. This technique employs homomorphic filtering and deconvolution to separate the voice from the speech which is assumed to be convolved.

Apart from the speech parameterization, another major challenge in voice conversion comes from the training techniques adopted. The pioneering work of [2] adopted vector quantization and spectrum mapping to convert the voice of a source to that of a target. Codebooks for spectrum parameters, power values, and pitch frequencies were created from training utterances. These mapping codebooks which are the representation of the correspondence between speakers' codebook is used as a weighting function for the conversion [2].

Several other techniques since then have been used. These include vector field smoothing [3], Gaussian mixture model [4], and Vocal tract length normalization [5]. A survey of literature shows that Gaussian mixture model is the most popular technique. This is because they are good at making a generalization of large sample distributions to form smooth approximations of different mixture densities. The GMM is a weighted sum of Gaussian densities represented by (2).

$$P(x|\lambda) = \sum_{i=1}^{M} w_i g(x|\mu_i, \Sigma_i) \quad (2)$$

where x is a D-dimensional feature vector, $w_i$ the mixture weights, M the number of mixtures and $g(x|\mu_i, \Sigma_i)$ the component Gaussian densities. $\mu_i$ and $\Sigma_i$ are the mean vector and the covariance matrix respectively.

Reference [6] carried out a comparative study of voice conversion with neural network and Gaussian mixture model. The study confirms that the neural network based voice conversion outperform the GMM based conversion in term of quality and intelligibility. References [7] and [8] also proposed different variants of neural network for voice conversion.

Despite the achievements that have been made in voice conversion over the years, Over-smoothing still remains a challenge especially in parametric speech synthesis. This occurs as a result of losses in the spectral details of the synthesized speech thus producing a muffled and unnatural speech that is too distant from the targeted speaker. Deficiencies in spectral representation of the speech signal play a major role in these losses. The non-linear mapping ability of neural network with line spectral frequency coefficients and linear prediction analysis coefficients was proposed in this work to investigate spectral losses in voice conversion and alleviate the problem in parametric voice conversion.

## II. FEATURE EXTRACTION

### A. Spectral Feature Extraction

Linear predictive coding (LPC) analysis was used to extract the features of the speech and model the acoustic space of both the source and target speaker. To accomplish this, the training speech signal was first sampled and pre-emphasized with a high pass filter to boost the energy of the high-frequency components of the speech signal. The speech samples were then divided into several frames or blocks of length 25 milliseconds with a time step of 5 milliseconds amounting to an overlap of 20 milliseconds and Gaussian windowed to prevent end effect due to discontinuity at the edges of the framed speech. Each of the frames was then auto-correlated. The autocorrelation function for discrete data which is also the correlation of a function with itself is given by (3) [9].

$$R_{xx}(\tau) = \lim_{N \to \infty} \frac{1}{N} \sum_{t=1}^{N} x(t)x(t+\tau) \quad (3)$$

where $x(t)$ is the windowed signal, $\tau$ the lag, $R_{xx}(\tau)$ the autocorrelation function and N the total number of samples.

The energy of the signal as shown in (4) is maximum where the lag equals zero.

$$R_{xx}(0) = \lim_{N \to \infty} \frac{1}{N} \sum_{t=1}^{N} x^2(t) \quad (4)$$

The system of Equations resulting from (4) yields a Toeplitz matrix which is solved recursively with the Levinson-Durbin's algorithm (also known as the Yule-Walker auto-recursive Equations) to yield the predictor filter coefficients that represent the vocal tract filter. A filter order of 24 was used. To ensure that the vectors have equal time alignment, time warping was done on the feature vectors to ensure they have the same duration before been mapped.

### B. Stabilization of LPC Filter Coefficients

According to [10], LPC coefficients are sensitive to quantization. Consequently, the predictor coefficients for both the source and target speaker were converted to line spectral pair using (5) and (6).

$$P(Z) = \prod_{k=1}^{p+1}(1 - e^{j\Omega_k Z^{-1}}) \quad (5)$$

$$Q(Z) = \prod_{k=1}^{p+1}(1 - e^{j\Theta_k Z^{-1}}) \quad (6)$$

where $\Omega$ and $\Theta$ are the line spectral frequencies (LSF), and $P(Z)$ and $Q(Z)$ the spectrum pair filter polynomial.

The same Neural Network architecture was used for the mapping of both the LPC and LSF coefficients and the result discussed.

### C. Extraction of Excitation Signal

The residual signal (also known as the excitation signal) carries the prosodic information (pitch) of the speaker. Equation (7) shows that the excitation signals which are also referred to as the error signal are the difference between the original signal and the predicted one.

$$e(n) = s(n) - \hat{s}(n) \quad (7)$$

where $e(n)$ is the error signal, $s(n)$ the speech signal and $\hat{s}(n)$ the predicted signal.

The excitation signal was obtained by filtering the input signal with the inverse transfer function of the linear prediction analysis filter. The linear predictor coefficients and the error signal or excitation signal form the acoustic vector space of each speaker.

### D. Neural Network Architecture

Artificial Neural Network is a family of models inspired by biological neural networks [11] and is designed to train, fit and validate data. The network is constituted by an organized set of layers that contains inter-connected nodes with an activation function.

The neural network usually is presented with a set of data known as training data. The network then adjusts its weights and biases in order to establish a pattern or relationship among the data. After adequate training, the network could estimate, classify and make predictions from new data based on the generalization it has established during training.

Complex problems like data classification and pattern recognition can be solved by combining multiple hidden layers to yield a multi-layer perceptron. A multi-layer feedforward network with back-propagation was used in this work to map the feature vector of the source into the acoustic space of the target. The justification for the configuration of the network chosen is shown in Table 1.

Result For Different Neural Network Configuration

TABLE I. RESULT FOR DIFFERENT NN CONFIGURATION

| NN Configuration | NN Training Results | |
|---|---|---|
| | *MCD (dB)* | *Training Duration (minutes)* |
| 24-30-24 | 1.1744 | 3.16 |
| 24-50-24 | 1.0288 | 6.20 |
| 24-25-25-24 | 1.8768 | 3.27 |
| 24-25-50-25-24 | 1.8015 | 12.50 |

*E. Result Evaluation*

The Mel-cepstral distortion (MCD) is seen as one of the best objective error measuring model. According to [12], it has good correlation with subjective test results. The MCD is a weighted Euclidean distance and has also been used by [13] and [14]. The Mel-cepstral distortion is given by (8).

$$mcd = \frac{10}{\ln(10)} \sqrt{2 \times \sum_{i=1}^{24}\left(mc_i^t - mc_i^p\right)^2} \quad (8)$$

where $mc_i^t$ and $mc_i^p$ are the acoustic vector of the i-th frame for the target and source speakers respectively. The smaller the MCD value, the better the conversion. A small value for the MCD indicates that there is only is very minimal spectral distance between the two vectors considered while a large value suggests otherwise.

## III. EXPERIMENTAL SETUP

One hundred phonetically balanced words were recorded and sampled at 11025 samples per second. Each of the words spans duration of 0.62 second. The words are framed and windowed in order to prepare them for linear prediction analysis. These words were pronounced by 4 different speakers, two female and two male within the age bracket of 25 and 28 years. Table 1 shows the average pitch of each of the four speakers used for this work and the full model employed in Fig 1.

TABLE II. PITCH INFORMATION OF SELECTED SPEAKERS

| Speaker | Speakers' Pitch | | |
|---|---|---|---|
| | *Minimum pitch (Hz)* | *Maximum Pitch (Hz)* | *Average Pitch (Hz)* |
| Male 1 | 96.54 | 134.82 | 120.86 |
| Male 2 | 84.45 | 118.30 | 102.89 |
| Female 1 | 213.94 | 267.93 | 245.68 |
| Female 2 | 192.33 | 259.03 | 226.32 |

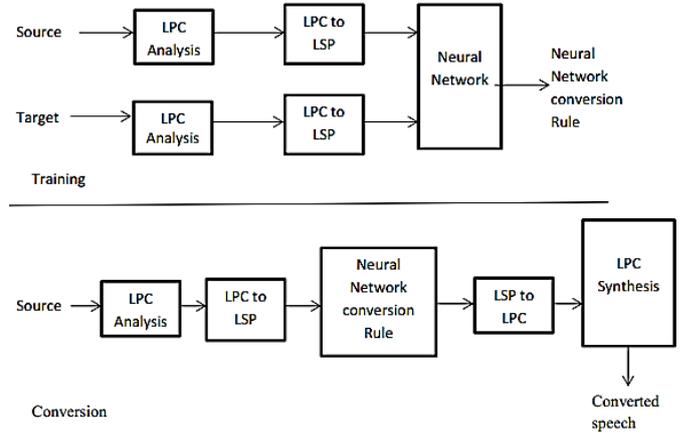

Fig. 1. Neural Network Conversion Model

The conversions were done using both the linear prediction analysis coefficients and the line spectral frequency coefficients with observations and results noted. Artificial neural network was used to map the LPC vector of the source to that of the target in order to investigate the stability of the LPC coefficients after training. The LPC vectors were converted into LSF vectors and mapped using the same neural network configuration to investigate the stability of the LPC poles and its effect on speech synthesis. The experimental setup was implemented for a male to male conversion, male to female conversion, female to male conversion and female to female conversion. Objective evaluation was carried out to evaluate the performance of the model in each of the four scenarios that were simulated using the MCD.

## IV. RESULT AND DISCUSSION

*A. Stability Of LPC and LSF Filter Coefficients*

The major cause of over-smoothing is the loss of spectral details. The LPC filter coefficients were found to be highly susceptible to this after training. Some of the filter poles as shown in Fig. 2 moved outside the unit circle. This is largely due to their sensitivity to quantization noise. The over-smoothing effect of this instability led to a poor conversion result. Fig. 3 shows that converting the LPC filter coefficients to LSF can improve the stability of the system. All the poles

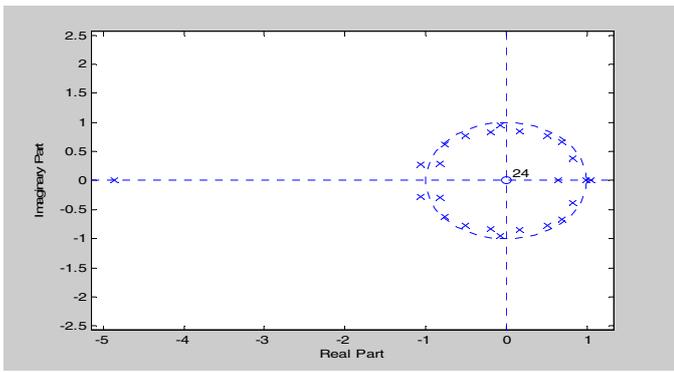

Fig. 2. Pole-Zero Plot For the LPC Filter Coefficients

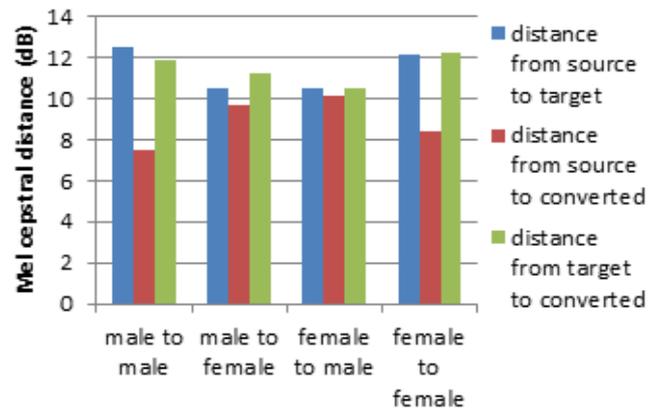

Fig. 4. Mel cepstral distance for Neural Network Mapping of LPC coefficients.

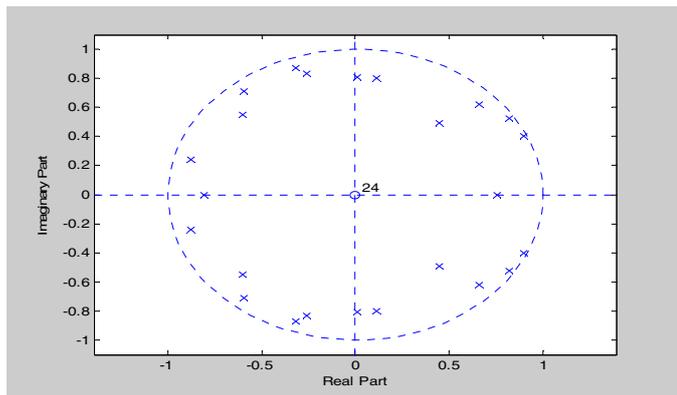

Fig. 3. Pole-Zero Plot For the LSF Filter Coefficients

### B. Neural Network Mapping Of LPC and LSF Filter Coefficient

are enclosed in the unit circle signifying an implementable system

Fig. 4 shows the Mel-Cepstral Distance from source speech to target speech, source speech to converted speech and target speech to converted speech for the LPC filter coefficient conversion. In all of the four cases of male to male conversion (M2M), male to female (M2F), female to male (F2M) and female to female (F2F), poor conversion result was obtained. Instead of a reduction in the spectral distance between the target and the converted, there was an increase of 37.3, 14.0, 3.5 and 31.6 percent in the M2M, M2F, F2M and F2F conversion respectively.

Fig. 5 is the result of the NN mapping of LSF filter coefficients. The result indicates the conversion went in the right direction for all cases. There was a spectral decrease of 88.4, 90.1, 88.6 and 88.4 percent for the M2M, M2F, F2M and F2F conversion respectively.

To investigate the performance of the model on noisy data, real life recorded utterances were further used to and the result for all the conversion type carried out is shown in Fig. 6. On the average, a 35.7 percent spectral reduction was obtained between the targeted and converted speech utterances. The model result was further compared with other voice conversion model tested with noisy data and the result shown in Table III.

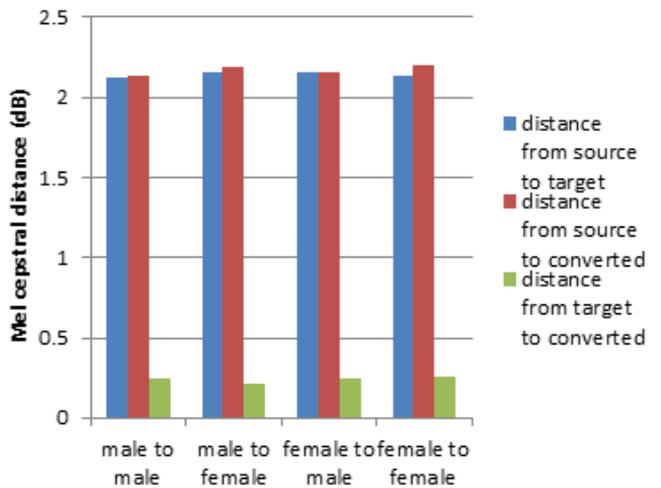

Fig. 5. Mel cepstral distance for Neural Network LSF Mapping

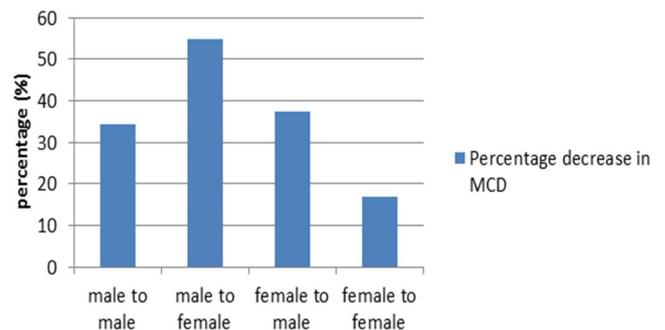

Fig. 6. Percentage decrease for noisy (live recorded) speech utterances

TABLE III. COMPARISON WITH OTHER MODELS

| Author | Model Comparison | |
|---|---|---|
| | *Model* | *Percentage of Spectral Decrease Between Target and Converted Speech* |
| Hashimoto and Higuchi (1995) | Vector Field Smoothing | 25.00 |
| Helander et al. (2010) | Gaussian Mixture Model with Partial Least Square Regression | 34.25 |
| Aihara et al. (2014) | Non-Negative Matrix Factorization | 6.81 |
| Developed Model | Neural Network with LSF Coefficient | 35.70 |

## V. CONCLUSION

This work uses LPC and LSF filter coefficient to underscore the impact of speech parameterization in voice conversion. Losses in the spectral details lead to over-smoothing, thus affecting the quality of the synthesized speech in term of its distance between the target and the converted speech. Neural network and LSF was used to carry out the conversion and result shows an improvement within the range of existing model.

ACKNOWLEDGMENT

Special thanks the members, staff and management of the Department of Electrical and Electronics Engineering, Federal University of Technology, Akure for their support and input into this work. Your efforts are well appreciated